\documentclass[a4paper]{jpconf}
\usepackage{graphicx}
\begin{document}
\title{Shell-model study of exotic Sn isotopes with a realistic effective interaction}

\author{A Covello$^{1,2}, \ $L Coraggio$^2$, A Gargano$^2$ and N Itaco$^{1,2}$}

\address{$^1$Dipartimento di Scienze Fisiche, Universit\`{a} di Napoli Federico II, \\
Complesso Universitario di Monte S. Angelo, I-80126 Napoli, Italy \\
$^2$Istituto Nazionale di Fisica Nucleare, \\
Complesso Universitario di Monte S. Angelo, I-80126 Napoli, Italy\\}

\ead{covello@na.infn.it}

\begin{abstract}
We report on a shell-model study of Sn isotopes beyond $N=82$ employing a realistic effective interaction derived from the CD-Bonn nucleon-nucleon potential renormalized through use of the $V_{\rm low-k}$ approach. At present, the most exotic Sn isotope for which some experimental information exists is $^{134}$Sn with an N/Z ratio of 1.68. It is the aim of our study to compare the results of our calculations with the available experimental data and to make predictions for the neighboring heavier isotopes which may be within reach of the next generation of radioactive ion beam facilities. The very good agreement between theory and experiment obtained for $^{134}$Sn gives confidence in the predictive power of our realistic shell-model calculations.

\end{abstract}

\section{Introduction}

The study of Sn isotopes beyond $N=82$ is currently the subject of great interest, as it allows to explore for possible changes of the shell-model structure when moving towards the neutron drip line.
At present, however, some experimental information is only available for $^{134}$Sn. Beyond this nucleus, which has an N/Z ratio of 1.68, lies a neutron-rich {\it terra incognita} which may become accessible to the next generation of radioactive ion beam facilities. In this perspective, it is challenging to make predictions which may stimulate, and provide guidance to, future experiments.
In fact, over the past several years various shell-model studies have been performed \cite{Coraggio02,Covello07,Kartamyshev07,Sarkar10} predicting spectroscopic properties of hitherto unknown neutron-rich Sn isotopes. In these studies both empirical and realistic effective interactions have been used, the latter being based on the CD-Bonn nucleon-nucleon ($NN$) potential model. In this connection, it should be mentioned that a main feature of the study of Ref. \cite{Covello07} is that the short-range repulsion of the bare potential is renormalized by use of the $V_{\rm low-k}$ approach \cite{Bogner02}. 

A peculiar property of $^{134}$Sn is the position of its first $2^+$ state which, lying at 726 keV excitation energy, is the lowest first-excited $2^+$ level observed in a semi-magic even-even nucleus over the whole chart of nuclides. This feature as well as the other observed excited states
- $4^+$, $6^+$, and $8^+$ - have been successfully reproduced by the realistic calculations of  \cite{Covello07}. In that study a very good agreement was also obtained with the experimental $B(E2;0^+ \rightarrow 2^+_1)$  value measured via Coulomb excitation of
a radioactive $^{134}$Sn beam  \cite{Beene04}.

Very recently, two sophisticated experimental studies \cite{Dworschak08,Jones10} have provided new information of great relevance to the study of Sn isotopes beyond $N=82$.
More precisely, in  \cite{Dworschak08} a high-precision Penning trap mass measurement has revealed a 0.5 MeV discrepancy with respect to previous $Q_\beta$ measurements. This finding provides clear evidence of the robustness of the $N=82$ shell closure.  In \cite{Jones10}, 
the magic nature of  $^{132}$Sn has been explored through the study of the single-particle states in 
$^{133}$Sn populated by a (d,p) reaction in inverse kinematics. This study has evidenced the purity 
of the single-neutron excitations in $^{133}$Sn and has identified a strong candidate for the
$2p_{1/2}$ single-particle state at 1.363 MeV excitation energy. This is about 300 keV lower
than the previous accepted value \cite{Hoff96}. 

Based on the above, we have found it interesting to revisit our previous realistic shell-model calculations for $^{134}$Sn and extend them to the heavier isotopes up to $A=140$.
In this paper we first give a brief description of the theoretical framework in which our calculations are performed and then present and discuss our results which are, for the most part, predictions for possible future experiments. A short summary is given in the last section.

\section{Outline of calculations}
We have performed shell-model calculations for even  tin isotopes beyond the $N=82$ shell closure,
namely, from   $^{134}$Sn to  $^{140}$Sn.  
We have taken doubly magic $^{132}$Sn as closed core and let the valence neutrons occupy the six levels 
$0h_{9/2}$, $1f_{7/2}$, $1f_{5/2}$, $2p_{3/2}$, $2p_{1/2}$,  and $0i_{13/2}$
of the $82-126$ shell. 
Our adopted single-particle (SP) energies relative to the $1f_{7/2}$ level are (in MeV):
$\epsilon_{ p_{3/2}}=0.854$, $\epsilon_{ p_{1/2}}=1.363$, $\epsilon_{h_{9/2}}=1.561$,
$\epsilon_{ f_{5/2}}=2.005$,
and $\epsilon_{ i_{13/2}}=2.694$. These values are taken from the experimental
spectrum of  $^{133}$Sn \cite{Jones10,NNDC}, with the exception  of the $0i_{13/2}$ level which was 
determined from the
position of the 2.434 MeV level in $^{134}$Sb assumed to be a
$10^+$ state of $\pi g_{7/2} \nu i_{13/2}$ nature. As for the $2p_{1/2}$ level, we use 
the energy reported in the  recent experiment of Ref. \cite{Jones10}. It is worth emphasizing
 that in this  experiment the spectroscopic factors  
of the four levels $1f_{7/2}$, $2p_{3/2}$, $2p_{1/2}$,  $1f_{5/2}$ were extracted, which 
provide a clear confirmation of  their SP nature.  
The absolute energy of the $1f_{7/2}$ level relative to $^{132}$Sn was placed
at $-2.371$ MeV, as resulting from the new mass measurement of $^{133}$Sn~\cite{Dworschak08}. 

The shell-model effective interaction has been derived within the framework of
perturbation theory~\cite{Coraggio09} starting, as mentioned in the Introduction,  from the 
CD-Bonn $NN$ potential ~\cite{Machleidt01} renormalized by way of the so-called $V_{\rm low-k}$ 
approach~\cite{Bogner02}.  
More precisely, we start by  deriving $V_{\rm low-k}$
with a cutoff  momentum of $\Lambda=2.2$ fm$^{-1}$. Then, using this  
potential plus the Coulomb force for protons, we calculate  the two-body matrix elements of 
the effective interaction  by means of the $\hat Q$-box folded-diagram expansion, with 
the $\hat Q$-box  including all diagrams up to second order  in the 
interaction.
These diagrams are computed within the harmonic-oscillator basis using intermediate states 
composed of all possible hole states and particle states restricted to the five proton and 
neutron shells above the Fermi surface. The oscillator parameter is 7.88 MeV for the $A=132$ region, 
as obtained from the expression  $\hbar \omega= 45 A^{-1/3} -25 A^{-2/3}$. 
The calculations have been performed by using the 
OSLO shell-model code \cite{Oslo}.

\section{Results}

\begin{figure}
\begin{center}
\includegraphics[width=15pc]{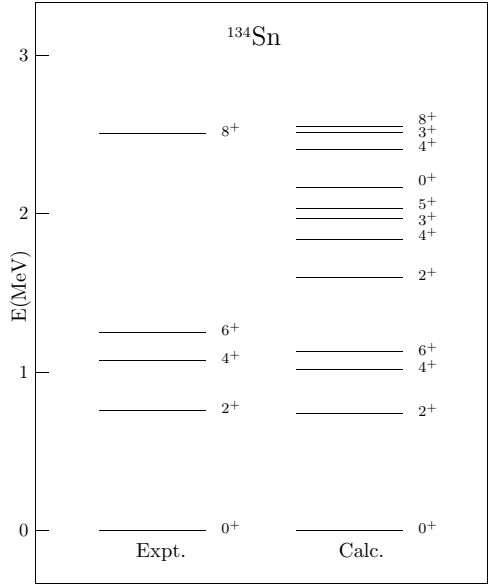}
\end{center}
\caption{\label{fig1} Experimental and calculated spectrum of $^{134}$Sn.}
\end{figure}

We start this section by discussing our results for  $^{134}$Sn,  the 
only tin isotope beyond $N=82$ for which some experimental information
is available. In figure~\ref{fig1} all the observed levels~\cite{NNDC} are 
reported and compared with the calculated ones up to about 2.5 MeV excitation energy.
The experimental spectrum consists of  a group of low-lying levels separated by an energy gap 
of about 1.5 MeV from the highest observed level which has $J^{\pi}=8^+$. The former  correspond 
to the four calculated members of the $\nu(f_{7/2})^2$  multiplet while the latter to the 
maximum-aligned  state of the $\nu f_{7/2}h_{9/2}$ configuration and all energies are 
very well reproduced by the theory.  In the energy interval between 1.1 and 2.5 MeV our calculation predicts the existence of other states, namely the 
members of the $\nu f_{7/2}p_{3/2}$ and $\nu f_{7/2}p_{1/2}$ multiplets,  and of
the $0^+$ state  of  the $\nu(p_{3/2})^2$ configuration. 
It is worth noting that in our previous calculation \cite{Covello07} the $3^{+}_{2}$ and $4^{+}_{3}$ 
states were predicted to lie above the $8^+$ state, their downshift in energy being directly related 
to the new position of the $p_{1/2}$ SP level.

We have also calculated some electromagnetic properties 
and our predicted values for  the first four excited states are reported in table~\ref{tab1}, 
where the two  experimental transition rates  presently  available~\cite{Beene04,NNDC} are also shown. Our calculations have 
been performed using an effective neutron charge of 0.7$e$~\cite{Coraggio02},  which leads to 
$B(E2)$ values for both  the $6^{+}\rightarrow 4^{+}$ and  
$2^{+}\rightarrow 0^{+}$ transitions quite close  to the experimental ones.

\begin{center}
\begin{table} [h]
\caption{\label{tab1} Electromagnetic transition rates (in W.u.), quadrupole (in efm$^2$) 
and magnetic  moments (in nm) in $^{134}$Sn.} 
\centering
\begin{tabular}{ccc}
\br
 &  Calc & Expt  \\
\mr
$B(E2; 2^{+}_{1} \rightarrow 0^{+}_{1})$ & 1.64 & $1.42 \pm 0.20$ \\
$B(E2; 4^{+}_{1} \rightarrow 2^{+}_{1})$ & 1.66 &   \\
$B(E2; 6^{+}_{1} \rightarrow 4^{+}_{1})$ & 0.82 & $0.89 \pm 0.17$    \\
$B(E2; 2^{+}_{2} \rightarrow 0^{+}_{1})$ & 0.35 &    \\
$B(E2; 2^{+}_{2} \rightarrow 2^{+}_{1})$ & 2.93 &    \\
$B(E2; 2^{+}_{2} \rightarrow 4^{+}_{1})$ & 0.23 &   \\
$B(M1; 2^{+}_{2} \rightarrow 2^{+}_{1})$ & 0.02 &  \\
$Q(2^{+}_{1})$ & -1.6 &   \\
$Q(2^{+}_{2})$ & -2.8 &  \\
$\mu(2^{+}_{1})$ & -0.57  &  \\
$\mu(2^{+}_{2})$ & -0.25 &  \\
\br
\end{tabular}
\end{table}
\end{center}

We hope that the above  predictions for $^{134}$Sn may provide guidance to future experiments aiming 
at the  identification of the missing states above the yrast $6^+$ state 
as well as at the   measurement of other properties,
which may certainly  give better insight into the structure of this exotic nucleus.

In this context, it should be mentioned that several interesting questions have been posed 
by the new, although still scarce, experimental data which have become available for some nuclei
in the $^{132}$Sn region. In particular,  special attention has been focused on the properties of the 
yrast $2^+$ states of tin and tellurium isotopes, whose excitation energy shows an asymmetric 
behavior with respect  to $N=82$. In fact, in $^{134}$Sn and $^{136}$Te  the $2^{+}_1$ energy undergoes 
a significant drop as compared to the $N=80$ isotopes, which was traced to a reduction
of the neutron pairing above the $N=82$ shell \cite{Terasaki02,Shimizu04}.   

With our realistic effective interaction we are able, as shown above, to reproduce with good accuracy 
the energy of  the yrast  $2^+$ state and its  $B(E2)$ transition rate in $^{134}$Sn. 
This  was also the  case for $^{136}$Te (see \cite{Covello07a}) whose $2^+$ state is described 
as the corresponding state in $^{134}$Sn with the two additional protons coupled to zero 
angular momentum. As a matter of fact, the $2^+$ state, as well as the other three lowest-lying states in $^{134}$Sn, are characterized by 
a very weak configuration mixing. Therefore the small $0^{+}-2^{+}$ spacing reflects
the small difference between the $<(f_{7/2})^{2}|V_{\rm eff}|(f_{7/2})^{2}>_{0^{+}}$ and 
$<(f_{7/2})^{2}|V_{\rm eff}|(f_{7/2})^{2}>_{2^{+}}$ matrix elements, which 
turns out to be  about 0.4 MeV. This result is due to the negligible contribution 
\cite{Coraggio09a} of 
the one particle-one hole  excitations in the effective interaction of two 
neutrons above the $N=82$ 
shell, which are instead responsible for the  increase of  pairing 
below this shell.  The weakness of the cross-shell interaction for $^{132}$Sn as compared with 
other lighter closed cores was also pointed out in Ref.~\cite{Kartamyshev07}, where it was noted  that the
excitation energy of the $3^-$ state in $^{132}$Sn is only 0.5 MeV smaller than the shell gap for
neutrons.

\begin{figure}[h]
\begin{center}
\includegraphics[width=20pc]{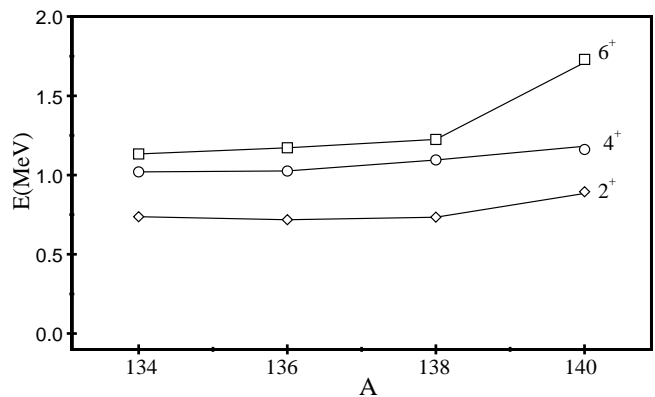}
\end{center}
\caption{\label{fig2}Calculated excitation energies of the yrast $2^+$, $4^+$,
and $6^+$ states in tin isotopes with $A=134$, 136, 138, and  140.}
\end{figure}

We have investigated the evolution of the yrast $2^+$ state and of the 
other $J \ne 0$ members of the  lowest-lying multiplet when adding neutron pairs. In figure ~\ref{fig2}
we report the excitation energies of the $2^{+}_{1}$, $4^{+}_{1}$,  and $6^{+}_{1}$ 
states as a function of A up to $^{140}$Sn. We see that the curves are almost flat, 
with a slight increase at $N=90$ for the $2^{+}_{1}$ and  $4^{+}_{1}$ which becomes more
significant  for the $6^{+}$ state.  

\begin{figure} [h]
\begin{center}
\includegraphics[width=12pc,angle=90]{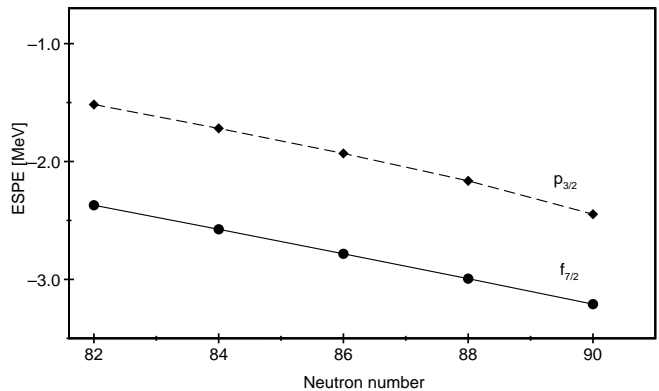}
\end{center}
\caption{\label{fig3}Effective single-particle energies of the $f_{7/2}$ and $p_{3/2}$ levels from $N=82$ to $N=90$.}
\end{figure}

This result is similar to that found  in \cite{Kartamyshev07}, the main difference being that our 
$4^{+}_{1}$  and $6^{+}_{1}$ curves lie about 100-300 MeV below those of \cite{Kartamyshev07}, 
while is in contrast to the conclusion of the recent work  of   \cite{Sarkar10}
claiming for a shell closure at $N=90$. The results of \cite{Sarkar10} predict an upshift of 
the $2^+$ state in $^{140}$Sn with respect to the lighter isotopes when using an empirical interaction (SMPN) as well as a realistic interaction with three-body monopole 
corrections (CWG3M).  This upshift  was related to the filling of the $f_{7/2}$ orbit at $N=90$
as a consequence of a significant increase in the difference  between the effective single-particle energies of the $f_{7/2}$ and 
$p_{3/2}$ orbits which, starting from 0.854 keV, becomes larger than 2 MeV at  $A=140$. 
As can be seen in figure~\ref{fig3}, our calculations do not predict such an increase. The
energy difference between the two orbits remains in fact almost constant and  even 
decreases slightly by  about 100 keV when moving from $N=82$ to $N=90$.
As regards the $6^+$ state, we find that its wave function in $^{140}$Sn is dominated by the 
$\nu(f_{7/2})^{6}(p_{3/2})^2$ configuration, which  explains its increasing in energy. 

In concluding this section, we discuss binding (relative to $^{132}$Sn) and 
one-neutron separation energies of the even Sn isotopes. The calculated values are reported in table~\ref{tab2}, where the experimental
available values for 
$^{134}$Sn are also shown. As mentioned in the Introduction, a new measurement of
the binding energy of $^{134}$Sn has revealed a 0.5 MeV deviation from the previous
observed value. With the new mass value the maximum of the 
neutron-shell gap for N=82 is at $Z=50$, instead of $Z=51$ as established by the 
old measurement.  We find an excellent agreement between theory and 
experiment, which shows the reliability of the $J=0$ matrix elements of our effective interaction. This confirms the weakening  of pairing for 
two neutrons beyond the $N=82$ shell. As regards the heavier  Sn isotopes, we see that the binding
energy keeps increasing  with increasing mass number, while the one-neutron 
separation energy remains practically constant.
Our calculations, in line with those of \cite{Kartamyshev07} and the estimations of 
mean field calculations,  seem to indicate that in this region we are still 
quite far from the neutron drip line.

\begin{center}
\begin{table} 
\caption{\label{tab2} Binding  energies and one-neutron separation energies (in MeV) in tin isotopes
with $A=134$, 136, 138, and  140. } 
\centering
\begin{tabular}{lcccc}
\br
& $^{134}$Sn &  $^{136}$Sn & $^{138}$Sn & $^{140}$Sn \\
\mr
BE {\small Calc} & 5.92 & 11.83& 17.68 & 23.41\\
BE {\small Expt}   & $5.916 \pm 0.150$  & & & \\
S$_n$ {\small Calc}  & 3.55 & 3.55  & 3.53 & 3.50 \\
S$_n$ {\small Expt}  & $3.545 \pm 0.152$ & & & \\
\br
\end{tabular}
\end{table}
\end{center}

\section{Summary}

We have presented here the results of a shell-model study of exotic even mass Sn isotopes from A= 134 to A= 140. In our calculations we have employed a realistic low-momentum effective interaction derived from the CD-Bonn $NN$ potential without using any adjustable parameter. 

The heaviest Sn isotope for which there is some, albeit scarce, experimental information is
$^{134}$Sn, so most of our results are  predictive in nature. We have shown that the available experimental data are very well reproduced by our calculations. This gives confidence in their predictive power, which we hope can be verified by experiment in the not too distant future.


\section*{References}
\begin{thebibliography}{99}
\bibitem{Coraggio02} Coraggio L, Covello A, Gargano A and Itaco N 2002 {\it Phys. Rev.} C {\bf 65} 051306(R)
\bibitem{Covello07}  Covello A, Coraggio L, Gargano A and Itaco N 2007 {\it Eur. Phys. J ST} {\bf 150} 93
\bibitem{Kartamyshev07} Kartamyshev M P, Engeland T, Hjorth-Jensen M and Osnes E 2007 {\it Phys. Rev.} C  {\bf 76} 024313 
\bibitem{Sarkar10} Sarkar S and Sarkar M S 2010 {\it Phys. Rev.} C {\bf 81} 064328 and references therein 
\bibitem{Bogner02} Bogner S, Kuo T T S, Coraggio L, Covello A and Itaco N 2002 {\it Phys. Rev.} C {\bf 65} 051301(R)
\bibitem{Beene04} Beene J R {\it et al.}  2004 {\it Nucl. Phys. } A {\bf 746} 471c
\bibitem{Dworschak08} Dworschak M {\it et al.} 2008 {\it Phys. Rev. Lett.} {\bf 100} 072501
\bibitem{Jones10} Jones K L {\it et al.} 2010 {\it Nature} {\bf 465} 454
\bibitem{Hoff96} Hoff P {\it et al.} {\it Phys. Rev. Lett.} {\bf 77} 1020
\bibitem{NNDC} Data extracted using the NNDC On-line Data Service from the ENSDF database, file revised as of October 4, 2010
\bibitem{Coraggio09} Coraggio L, Covello A, Gargano A, Itaco N and Kuo T T S 2009 {\it  Prog. Part. Nucl. Phys.} 
 {\bf 62} 135 and references therein
\bibitem{Machleidt01} Machleidt R 2001 \emph{Phys. Rev.} C \textbf{63} 024001
\bibitem{Oslo} Engeland T the Oslo shell-model code 1991-2006, unpublished
\bibitem{Terasaki02} Terasaki J, Engel J, Nazarewicz W and Stoitsov M, 2002 {\it  Phys. Rev.} C {\bf 66} 054313 
\bibitem{Shimizu04} Shimizu N, Otsuka T, Mizusaki T and Honma M 2004 {\it  Phys. Rev.} C {\bf 70} 054313 
\bibitem{Covello07a} Covello A,  Coraggio L, Gargano A and Itaco N 2007 {\it Prog. Part. Nucl. Phys.} {\bf 59} 401
\bibitem{Coraggio09a} Coraggio L, Covello A, Gargano A and Itaco N 2009 {\it Phys. Rev.} C {\bf 80} 021305(R)
\end{thebibliography}
\end{document}